# Virtual Therapy Exergame for Upper Extremity Rehabilitation Using Smart Wearable Sensors

Lauren Baron[1,*], Vuthea Chheang[1,*], Amit Chaudhari[2], Arooj Liaqat[1], Aishwarya Chandrasekaran[1],
Yufan Wang[1], Joshua Cashaback[3], Erik Thostenson[2], Roghayeh Leila Barmaki[1]

[1]*Department of Computer and Information Sciences, University of Delaware*, Newark, DE, USA
[2]*Center for Composite Materials, Mechanical Engineering, University of Delaware*, Newark, DE, USA
[3]*Department of Biomedical Engineering, University of Delaware*, Newark, DE, USA

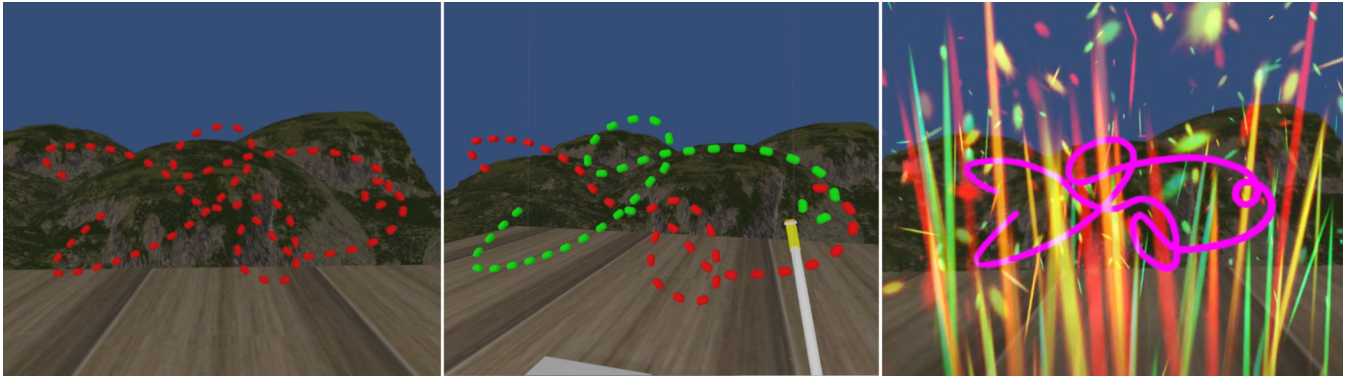

Figure 1: The immersive VR therapy exergame for upper extremity rehabilitation: (left) the virtual content used for the game, (center) the drawing exercise task with a virtual brush, and (right) the celebration effects after task completion.

## ABSTRACT

Virtual Reality (VR) has been utilized for several applications and has shown great potential for rehabilitation, especially for home therapy. However, these systems solely rely on information from VR hand controllers, which do not fully capture the individual movement of the joints. In this paper, we propose a creative VR therapy exergame for upper extremity rehabilitation using multi-dimensional reaching tasks while simultaneously capturing hand movement from the VR controllers and elbow joint movement from a flexible carbon nanotube sleeve. We conducted a preliminary study with non-clinical participants (n = 12, 7 F). In a 2 x 2 within-subjects study (*orientation (vertical, horizontal)* x *configuration (flat, curved)*), we evaluated the effectiveness and enjoyment of the exergame in different study conditions. The results show that there was a statistically significant difference in terms of task completion time between the two orientations. However, no significant differences were found in the number of mistakes in both orientation and configuration of the virtual exergame. This can lead to customizing therapy while maintaining the same level of intensity. That is, if a patient has restricted lower limb mobility and requires to be seated, they can use the orientations interchangeably. The results of resistance change generated from the carbon nanotube sleeve revealed that the flat configuration in the vertical orientation induced more elbow stretches than the other conditions. Finally, we reported the subjective measures based on questionnaires for usability and user experience in different study conditions. In conclusion, the proposed VR exergame has the potential as a multimodal sensory tool for personalized upper extremity home-based therapy and telerehabilitation.

---

[*]These authors contributed equally to this work.

# 1 INTRODUCTION

Physical therapy (PT) is a well-known treatment for effective rehabilitation. Among post-stroke survivors, as the potential target users for our research, patients often have musculoskeletal conditions, especially upper extremity functional limitations [29]. Approximately 80% of people who suffered a stroke experience motor impairments, including in their upper limbs [13]. Additionally, low adherence to PT exercises has been constantly reported, such as lack of motivation, slow recovery progress, and absence of mental support [1, 25, 28]. As a result, there is a pressing need to make conventional rehabilitation for upper limb mobility more interactive, engaging, flexible, and effective.

Virtual reality (VR) has shown immense potential to provide an engaging, entertaining, and enjoyable experience of PT exercises. Many studies have shown that VR is beneficial and preferred for rehabilitation in various ways, e.g., portability for home therapy, engaging virtual environment, and independence from distractions [12, 37, 38, 41]. While VR has demonstrated a lot of potential for PT, research on creative virtual therapy and task-oriented therapeutic exercises (e.g., exergames), especially for upper extremity rehabilitation, is still underrepresented [3, 12, 21]. In addition, a therapeutic assessment for PT in VR is needed. Most systems for virtual therapy only rely on information captured from VR controllers and hand tracking. Using VR trackers could provide the position of the hand/wrist joint, however, they do not fully capture the movement details from each individual joint. Our main contribution to the project is that we use a smart wearable sensor to provide a more accurate and quantifiable assessment of movement during gameplay.

In this work, a creative VR exergame is proposed for upper extremity therapy. We developed multi-dimensional reaching tasks and used a smart fabric-based carbon nanotube sensor on the elbow to capture limb movement from the arm and hands. A preliminary study *(n = 12, 7 F)* was conducted to evaluate the effectiveness and enjoyment of the proposed VR exergame in different model orientations and configurations. The objective and subjective measures such as task completion time, number of mistakes, the resistance change of the elbow sleeve sensor, and subjective questionnaires were assessed. Our research questions include the following:

- **RQ1**: How do model configuration and orientation influence the *therapeutic experience* in the VR therapy exergame?
- **RQ2**: How do model configuration and orientation influence the *electrical resistance changes* from the smart wearable sensor for upper extremity rehabilitation?
- **RQ3**: How is the subjective perception of the VR therapy exergame, such as *easiness, comfort, and enjoyment*, associated with different VR model conditions?

# 2 RELATED WORK

In this section, we report prior research related to the use of VR in rehabilitation and how the virtual content affects the overall user experience. We also describe wearable sensors used for VR therapy in the following sections.

## 2.1 VR Applications for Physical Therapy

VR can be used for training, decision-making, and rehabilitation in the physical therapy domain. For instance, Hartstein et al. [11] assessed the perceived ease of use and usefulness of VR learning experiences to promote the clinical decision-making of PT students.

Other works have also shown the potential of VR on upper limb rehabilitation for patients with stroke or Parkinson's Disease [2, 3, 31, 33]. Despite the well-known effectiveness of PT interventions for rehabilitation, various limitations, including time commitment, the intensity of labor and resources, dependability on patient compliance, geographical availability of special facilities, and costs/insurance coverage, have been reported [12, 18, 35]. Phenal et al. [32] explored the use of VR for children with upper limb motor impairment undergoing painful therapeutic processes within a hospital environment. In this study, they found that VR has the potential to improve functional disabilities, alleviate perceived pain, reduce the perceived difficulty of rehabilitation exercises, increase exercise duration and produce positive emotions toward the therapy.

In a systematic review of immersive VR for older adults, Campo Prieto et al. [5] provided preliminary evidence supporting immersive VR technologies' application in older adult populations. Multiple reviews on the efficacy of VR therapy conclude that the current evidence on the effectiveness of using VR in the rehabilitation of upper limb mobility in patients with stroke is limited and emphasised the need for more studies to support this effect while investigating different intervention types through rigorous studies [3, 12, 19, 36]. Xu et al. [40] developed a depth camera–based, task-specific VR game called *Stomp Joy* with an aim to assess its feasibility and clinical efficacy for post-stroke rehabilitation of the lower extremities, which was tested in a recent study. In contrast, our study is a depth-based therapeutic exergame for upper extremity rehabilitation.

## 2.2 Effects of Objective Content Adjustments on Virtual Experience

Mason et al.[24] explore the intersection between reaching tasks and depth analysis. Their study looks at reaching both physical and virtual targets with VR. They measured task completion time and wrist movements to determine how haptic and visual feedback influence the reaching movements. They also studied depth analysis for reaching kinematics and found that participants took more time decelerating towards smaller targets with haptic feedback provided. However, when haptic feedback was absent, deceleration time was constant. These findings suggest that virtual visual feedback for the moving limb and haptic feedback about contacting objects are important for performance in a virtual environment (VE). Without feedback, Fitts's law [23] to predict human movement does not always hold. They also supported that reaching tasks can be used for depth analysis in the VE.

In another study [10], participants had to reach real-world objects but were given different visual depth cues and only some used a VR headset. Interestingly, participants using the VR headset performed better and visual depth cues only had a minor impact on reaching performance. This shows how an immersive VR environment can be useful for reaching and depth perception. It also indicates that no one visual depth cue weighs more than others and a combination of many depth cues does not necessarily correlate to accuracy. While we measure depth perception objectively via task-completion time and subjectively via questionnaires, our goal is to incorporate Fitt's Law into our future studies to precisely measure reaching task-completion time in simpler models.

In another study, Gagnon et al. [9] aimed to assess whether feedback from reaching improves depth judgment and if re-calibration changed due to feedback across reaching behaviors. They tested judgments of action capabilities within a VE for two different reaching behaviors, reaching out and reaching up. Only some participants received feedback on whether they reached the target dot. They found that reach was initially overestimated, but over feedback blocks, perceptual estimates decreased and became more accurate. They also found that targets just beyond reach were more difficult to judge. Feedback on reaching activities on objects placed far away in an immersive VE is critical for improving depth perception.

In a study about depth perception in an immersive VE, 3D objects and closer objects were found to significantly provide better and more accurate depth estimation than 2D objects and far away objects [27]. When looking at hitting a distant target in a VE, participants had to use greater upper limb motor functions like muscular effort and torque of the shoulder to hit more distant objects [30]. VEs can be used to assess estimates of action capabilities and improve those estimates through visual-motor feedback. Reaching tasks, especially for distant and flat objects in a VE, require more upper limb effort and are harder to judge distance/depth. When someone overextends their elbows, rigid joints, etc., their different formations and movements can affect depth cues and impressions. Our study is one of the kinds that attempts to take elbow movements via resistance changes from a flexible nanotube wearable sensor into account in VE reaching tasks.

Kioumourtzoglou et al. [17] studied how extending the forearm with the elbow at 80 degrees can be used to measure a sense of kinesthesis and how we perceive our body's movement. Palaniappan et al. [30] used inverse kinematics to analyze joint angle positions, joint reaction forces, and joint torque to show how VR therapy is more effective than conventional therapy for rehabilitation. The precision of depth and directional judgment is affected by the pendular (contracting and relaxing muscles) motion of the limb segments.

## 2.3 Wearable Sensors for Virtual Therapy

The usability of wearable sensors during VR-based PT is increasingly becoming popular due to their accessibility. As it allows the monitoring of the quantity and quality of body movement, the data collected from smart wearables can provide effective measurements, thus better treatments for patients in the process of rehabilitation. Brandão et al. [4] explore the feasibility of unsupervised physical therapy for rehabilitation at patients' homes. This study was conducted on patients with hemiparesis due to stroke. The patients were trained with an inertial measurement unit (IMU) in VR. The participants were able to use the system in their homes without any supervision. It was found that the arm function of these patients improved significantly. Moreover, the participants were able to complete each session of rehabilitative therapy only in six weeks.

VR-based techniques have been used in healthcare to provide more accessible rehabilitation options to patients with disabilities. However, VR combined with wearable devices has made this experience of recovery more pleasant and valuable for patients [14, 39]. VR provides patients with an immersive experience in a virtual world and gives them the capability to interact with virtual objects using motion sensors. This attribute has made VR-based rehabilitation a promising tool to promote the active participation of patients in their therapy process and produce better motor recovery [16]. The data collected from wearable sleeves while performing VR physical therapy can be visualized and interpreted by therapists, patients, and caregivers. Even if they do not have any technical knowledge, an overview of patient performance for each session was found to be effective for them [26].

In another study, Lee et al. [20] examined the integration of wearable sleeves in VR-based physical therapy. In this study, the data was collected from wearable sensors in a VR-based goal-directed shoulder rehabilitation system to analyze task performance and improvement after each training session. The study was conducted on patients with frozen shoulders where they performed shoulder muscle strengthening, and core muscle strengthening exercises. While patients were performing exercises by interacting with the VR environment, the sensors were secured to their shoulders to measure the range of motion (ROM). The data collected from training sessions suggested that the usage of wearable sensors in VR therapy can provide better information to offer customized individual training programs. Hence, it is suggested that rehabilitative games for at-home VR therapy using wearable sensors are feasible and safe despite the lack of supervision [39].

## 3 MATERIALS AND METHODS

In the following sections, we describe the participants, apparatus, study procedure, study design with dependent and independent variables, and hypotheses of the user study.

### 3.1 Participants

For the experiment, a priori power analysis was conducted to determine the sample size for interaction effects for ANOVA (repeated measures, within factors) F tests. We used G*Power for a large effect size $\eta_p^2$ : 0.14 which gave the effect size f = 0.403 [8]. With a power of 0.80, one group, and four measurements, the result was a total sample size of 10. To keep a balanced design of three participants for each of the four measurements while still being greater than *n = 10*, we recruited a participant pool of *12* non-clinical volunteers (*5* males and *7* females; age ranged from *20 - 29, M=22.67, SD=2.78*). There was no monetary compensation for participation.

Eight out of *12* participants (*66.66%*) had Asian, or Pacific Islander ethnicity, and four (*33.33%*) had a Caucasian or White ethnicity. Most participants had prior VR but lacked prior video game experience; ten (*83.33%*) had used VR headsets before, but only three (*25%*) reported playing video games daily or weekly. Two participants (16.67%) have previously experienced a severe upper-body injury, either due to sports or other incidents, and needed to participate in rehabilitation sessions to recover. Only one participant (8.33%) was visually impaired beyond their glasses/contact lenses.

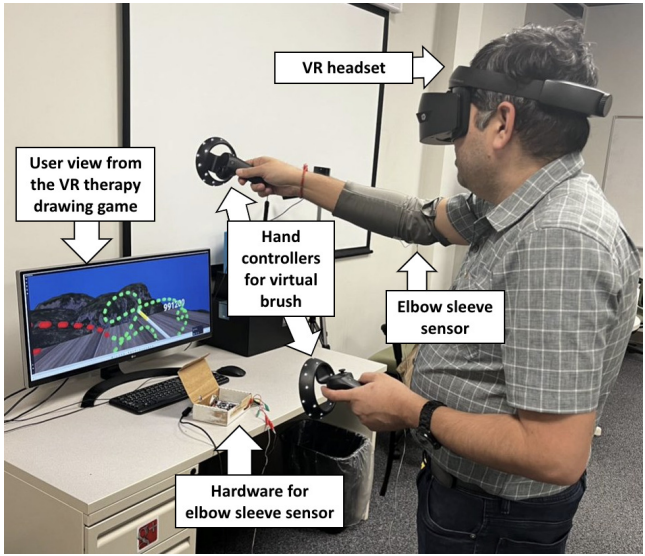

**Figure 2: The VR Therapy study setup for participants.**

## 3.2 Apparatus

An interactive VR drawing exergame was developed using the Unity game engine (version *2019.1.0f2*). In the drawing exergame, participants were presented with a welcoming screen in which they could choose drawing contents: the flat fish and curved fish. The models were developed using licensed Autodesk Maya and Blender.

The virtual environment was designed with a simple, serene mountainscape, a blue sky, and a wood floor to identify the action space. The only objects in the user's personal space were the virtual representations of the hand controllers (a cube for the non-dominant hand that controls the model's position/rotation and a paintbrush for the user's dominant hand to perform the task). This virtual environment allowed users to focus on the task in a relaxing, distraction-free environment without any other cues that could disturb the user's depth perception.

The primary task of users in this exergame was to connect the dots of a traceable outline of the fish model(s) using the paintbrush. When each dot was hit, it turned from red to green, and a positive audio feedback sound was played to the user. When all the dots were green, meaning the user successfully connected all the dots of the drawing, they were celebrated by visual firework animations with sound effects. This positive audio-visual feedback can offer a more guided, targeted reaching, and therapeutic experience to users as shown in preliminary research on the proper depth judgment via feedback cues [9].

A fabric-based carbon nanotube sensor [7] was used on the elbow, and the electrical resistance change of the sensor was recorded for elbow flexion/extension throughout the drawing task. The participants were asked to use their dominant hand for drawing and wearing the elbow sleeve (see Figure 2). The other controller was used to adjust the dotted model to the height or position the user felt most comfortable with.

## 3.3 Study Procedure

Once participants arrived, we received verbal consent after describing the purpose of our study, their responsibilities, and their right to withdraw and take breaks when needed. After consent, the participants were randomly assigned the order to conduct the four study conditions based on the *2 × 2* study design. Participants then filled out a pre-questionnaire about their demographics, prior experience with VR/video games, experience with physical therapy/exercise, and experience with visual impairments. After getting an explanation of study directions, participants walked to the middle of the room to stand in a cleared area. We then adjusted the elbow sleeve sensor to be right on top of their dominant hand's elbow, gave them the headset to put on, and placed their hand controllers so they can draw with their dominant hand. Their first session was one of the following conditions: flat fish vertical, flat fish horizontal, curved fish vertical, or curved fish horizontal.

When starting the task, we triggered the data collection scripts to record the objective measures. Once the task was completed and the victory animation had finished playing, the participants were asked to take off the VR headset and controllers and complete the post-questionnaire for the first session they had just experienced. The participants were then asked to put on the VR headset again and use the VR hand controllers to do their next task in a new configuration and orientation (the order of sessions was based on the *4 × 4* balanced Latin square). Once again, participants fill out the post-questionnaire for each of the new configurations/orientations. The entire experimental session for each participant took 20-30 minutes. A unique ID was generated by each participant and was repeatedly used in the completion of the questionnaires and saving of the data files to keep track of their data while preserving their anonymity.

## 3.4 Study Design

We had a *2 × 2* within-subjects study with a balanced design with two factors: (1) *orientation*: horizontal vs. vertical, and (2) *configuration*: flat vs. curved. Thus, four study conditions included flat vertical, flat horizontal, curved vertical, and curved horizontal for the fish model shown in Figures 3 and 4. The order to complete these four conditions was based on a balanced Latin Square (*4 × 4*) [15] to mitigate the learning effects of win-subjects design. There were no trials in the studies because there were no identical conditions.

We collected both objective and subjective measures to assess the perceived difficulty of the drawing performance and other (depth) perceptions related to the VR therapy experience through quantitative data collection and questionnaires. In the following, objective measures including independent and dependent variables, and subjective measures from the questionnaire are described.

*3.4.1 Independent Variables.* The user study was planned as within-subject design with a two-factor test. The two factors were defined by two independent variables: *orientation* and *configuration*.

For *orientation*, the drawing contents were rotated around the x-axis for a vertical view of the model and a horizontal view of the model (see Figure 3). The difficulty levels of these orientations were empirically tested while preparing the experiment.

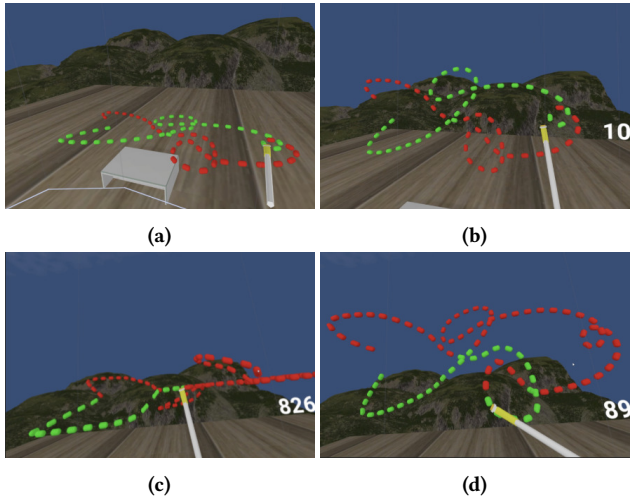

Figure 3: Four different conditions of the study: two orientations and two configurations of the fish model: (a) horizontal orientation, flat configuration, (b) vertical orientation, flat configuration, (c) horizontal orientation, curved configuration, and (d) vertical orientation, curved configuration.

- **Horizontal**: Participants performed the drawing activity while looking at the model facedown, with their reaching motions moving primarily out and in.
- **Vertical**: Participants performed the drawing activity while looking at the model head-on/straight up, with their reaching motions moving primarily up and down.

For *configuration*, two virtual drawing contents were objectively adjusted with different numbers of drawing dots and with different dimensions and depths (see Figure 3 and Figure 4). The difficulty levels of these virtual contents were empirically tested while preparing the experiment.

- **Flat**: The drawing content was the shape of an abstract outline fish image with 69 drawing dots, which could be easier to draw because they are all on the same z-plane.
- **Curved**: The drawing content was the shape of an abstract outline fish image with 91 drawing dots, which could be harder to draw because some dots are closer/farther than others with respect to the z-axis and more dots were required to add this dimension.

*3.4.2 Dependent Variables.* Three core measurements to objectively evaluate the drawing performance was defined as dependent variables.

- **Normalized Task Completion Time (TCT)**: The completion time of the drawing task was calculated based on the starting time when the participant hits the first dot and the ending time when the final dot of the model is hit. We normalized TCT over the number of drawing dots for a fair comparison between model configurations. Therefore, the measuring unit for TCT was *task completion time per drawing dot–still in seconds*.

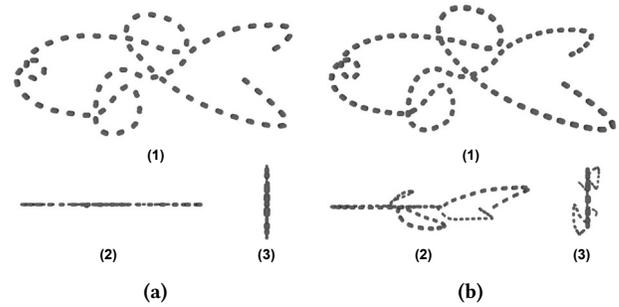

Figure 4: Detailed views from two conditions of the VR Therapy exergame: (a) vertical flat fish model and (b) vertical curved fish model, with the following view directions: (1) side view, (2) top-down view, and (3) front view.

- **Normalized Number of Mistakes**: The number of mistakes, e.g., when the participants missed any drawing dots while performing the tasks, was logged during the study. We normalized the number of mistakes over the TCT. Thus, the measuring unit was the *number of mistakes per second*.
- **Normalized Resistance Change**: The sensor's electrical resistance over the elbow changes during flexion and extension. Negative resistance change values were ignored and considered outliers. The percentage resistance change was calculated according to Equation 1, where $R$ is resistance at stretch and $R_m$ is minimum resistance value at no stretch [7]. We calculated the mean value of percentage resistance change with identical experimental values.

$$ResistanceChange(\%) = ((R - R_m) * 100)/R_m \quad (1)$$

In addition, the resistance change data was normalized over the TCT for the fair compression among the participants. Hence, the measuring unit for resistance change is the *percentage of change per second*.

*3.4.3 Questionnaire.* For the subjective measures, we collected data on usability and perception based on subjective questionnaires, using Qualtrics survey platform. The questionnaire was adapted with a standardized questionnaire to evaluate the user experience in the immersive environment [34]. The questions were in five-point Likert scale (*1: strongly disagree, 5: strongly agree*), except for the "Willingness to Recommend," which was collected separately in a semantic differential scale (*0: Not at all likely, 10: Extremely likely*). Below we describe the questions for each of the subjective measures:

- *Easiness*: "It was easy to complete the virtual drawing task."
- *Comfort*: "I felt comfortable while completing the task."
- *Enjoyment*: "I enjoyed playing the creative drawing game."
- *Body stretch*: "Using the VR drawing activity, I stretched my arm out more than I normally do."
- *Depth perception*: "I could easily reach the objects and judge the distance from the objects in the creative drawing game."
- *Visual cues*: "The virtual drawing model was realistic" and "I could easily see things and objects in the creative drawing game."

- *Willingness to recommend*: "I would recommend this creative therapy game to friends or family members as an upper-limb therapeutic exercise."

We also collected the participants' general feedback through text entries, asking for thoughts on how to improve this activity for future use and their preferences for the study conditions.

### 3.5 Hypotheses

The hypotheses for this user study arose from the specified tasks and the therapy experience, resulting in the following:

- **H1**: Participants' objective performance on the VR therapy exergame will be influenced by the model *configurations*.
- **H1-1**: The curved model configuration (more drawing dots and dimensions) will produce more *electrical resistance changes*, compared to the flat configuration.
- **H1-2**: Participants' performance and experience will be improved with the flat configuration (the distance from the user to the dots is the same for each dot), compared to the curved configuration.
- **H2**: Participants' objective performance on the VR Therapy exergame will be influenced by model *orientation*.
- **H2-1**: The horizontal orientation (reaching out) will produce more *electrical resistance changes* compared to the vertical orientation (reaching up).
- **H2-2**: Participants' performance and experience will be improved with the vertical orientation compared to the horizontal orientation because they can better see the content's full shape.

## 4 RESULTS

We used *RStudio* with *R* for statistical computing. An analysis of variance (*ANOVA*) was used for data analysis with three dependent variables: *task completion time*, *number of mistakes*, and *resistance change*. In addition, we run a normality test for data normal distribution. We further analyzed the data with pairwise *t-tests* and the *Bonferroni* adjustment method to identify the differences between the conditions. The questionnaire results were analyzed descriptively.

In the following, we describe the results of statistical analysis, questionnaire results, and general feedback.

### 4.1 Statistical Analysis

The summary of the descriptive results for dependent variables is listed in Table 1 and shown in Figure 5. Moreover, the results of the statistical analyses with *ANOVAs* are listed in Table 2.

*4.1.1 Normalized Task Completion Time (TCT).* Statistically significant difference was found in the interaction effect between the *orientation* and *configuration* ($p < 0.03$). However, there were no significant differences in the main effect of the conditions. We further analyzed the data with the pairwise *t-test*, and we found a difference in the flat configuration between the *horizontal* and *vertical orientation* ($t = 2.49$, $df = 11$, $p < 0.03$) for TCT. The results show that vertical orientation was performed faster than the horizontal ones in the flat model configuration. However, this difference in the curved configuration is not significant.

*4.1.2 Normalized Number of Mistakes.* For the number of mistakes for each virtual content/model, we found no statistically significant differences between the orientation and configuration conditions. Trade to subjective results shows that the number of mistakes in flat configuration is averagely lower than in the curved model configuration. Furthermore, the number of mistakes in the flat model in the vertical orientation is relatively lower than the horizontal orientation. However, the differences are small between curved configuration.

*4.1.3 Normalized Resistance Change.* The resistance change is indicated by how much the elbow sleeve stretches. We found statistically significant differences in the orientation ($p < 0.04$), configuration ($p < 0.03$), and their interaction effect ($p < 0.01$) for *resistance change*. The results of the pairwise t-test show a significant difference between the *curved* and *flat* configurations ($t = -3.51$, $df = 23$, $p < 0.002$). In addition, a significant difference was also found in the vertical orientation between the *curved* and *flat* configurations ($t = -3.59$, $df = 11$, $p < 0.004$). For orientation, there was a difference between the *horizontal* and *vertical* orientation in the *flat* configuration ($t = -2.83$, $df = 11$, $p < 0.01$). The results indicate that the *flat* model induced more *elbow stretches* than the curved model in the *vertical* orientation. Furthermore, the *vertical* orientation induced more *elbow stretches* compared to the *horizontal* orientation during the drawing performance of the virtual content.

### 4.2 Questionnaire Results

The questionnaire results are shown in Figure 6. The questionnaire data was analyzed descriptively, and we describe the results for each of subjective measures as follows:

- **Easiness** The average scores of easiness are $M = 4.29$, $SD = 1.08$ for horizontal and $M = 4.21$, $SD = 1.10$ for vertical orientations. The perceived easiness of the virtual content for flat model was averagely higher than the curved model on both orientations.
- **Comfort** The comfort average score for horizontal ($M = 4.83$, $SD = 0.48$) is higher than the vertical ($M = 4.71$, $SD = 0.55$) version. The perceived comfort of the virtual content for flat model was averagely higher than the curved model for both orientations, but it was a small difference.
- **Enjoyment** The mean score of enjoyment for horizontal ($M = 4.29$, $SD = 1.27$) is averagely lower than the vertical ($M = 4.46$, $SD = 0.83$) orientation.
- **Body Stretch** The average score of body stretch show the horizontal ($M = 2.88$, $SD = 1.62$) and vertical ($M = 2.58$, $SD = 1.56$). The perceived body stretch while drawing the curved model was averagely more than the flat model for both orientations.
- **Depth Perception** The participants rated the depth perception for horizontal ($M = 4.08$, $SD = 1.18$) and vertical ($M = 4.25$, $SD = 1.15$). The perception of depth for flat virtual content was averagely better than for curved content for both orientations. The perceived easiness to reach objects and judge the distance from objects for participants was significant between versions ($p < 0.05$). When looking at how the versions compared in pairs for the depth perception scores, there was significance between the flat vertical content and

Table 1: Summary of descriptive results for drawing performance.

| Variable | Norm. Task Completion Time (s) | Norm. Number of Mistakes | Norm. Resistance Change (%) |
|---|---|---|---|
| Horizontal | 0.71 (0.43) [0.08] | 0.086 (0.09) [0.01] | 0.60 (0.47) [0.09] |
| Curved | 0.59 (0.32) [0.09] | 0.089 (0.07) [0.02] | 0.51 (0.41) [0.11] |
| Flat | 0.82 (0.51) [0.14] | 0.083 (0.11) [0.03] | 0.70 (0.52) [0.15] |
| Vertical | 0.61 (0.37) [0.07] | 0.066 (0.07) [0.01] | 0.89 (0.86) [0.17] |
| Curved | 0.72 (0.34) [0.10] | 0.088 (0.08) [0.02] | 0.45 (0.50) [0.14] |
| Flat | 0.50 (0.39) [0.11] | 0.044 (0.04) [0.01] | 1.34 (0.93) [0.27] |

*All entities are in the format: mean value (standard deviation) [standard error]. (Norm. Normalized)*

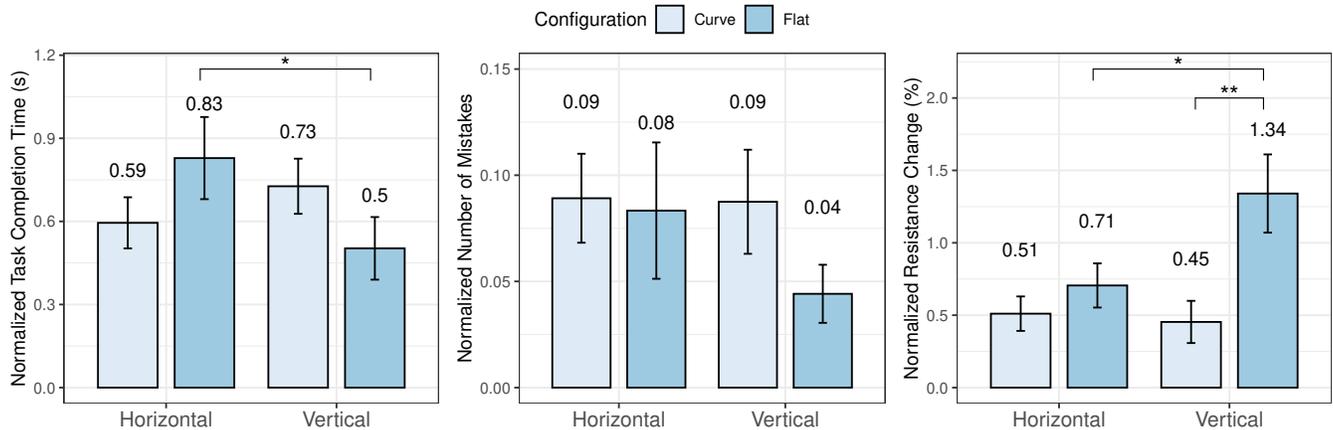

Figure 5: Results of dependent variables: (left) task completion time, (middle) mistakes, and (right) resistance change. Starred brackets mark significantly different conditions.

Table 2: Summary of statistical results with significance level and effect sizes ($p < .05$).

| Variable | df | F | p | Sig | $\eta^2$ |
|---|---|---|---|---|---|
| **Norm. Task Completion Time** | | | | | |
| Orientation | 1 | 1.292 | 0.27 | | 0.015 (small) |
| Configuration | 1 | 0.001 | 0.96 | | 0.00003 (small) |
| Orientation * Configuration | 1 | 5.872 | <0.03 | * | 0.082 (medium) |
| **Norm. Number of Mistakes** | | | | | |
| Orientation | 1 | 1.681 | 0.22 | | 0.016 (small) |
| Configuration | 1 | 2.068 | 0.17 | | 0.023 (small) |
| Orientation * Configuration | 1 | 0.479 | 0.50 | | 0.013 (small) |
| **Norm. Resistance Change** | | | | | |
| Orientation | 1 | 4.990 | <0.04 | * | 0.054 (medium) |
| Configuration | 1 | 11.823 | <0.005 | * | 0.168 (large) |
| Orientation * Configuration | 1 | 8.528 | <0.01 | * | 0.076 (medium) |

curved horizontal ($p < 0.02$). There was also significance between the flat vertical content and the curved vertical ($p < 0.03$).

- **Visual Cues** To get an insight into how accurate the visual cues were for the virtual contents; we measured the perceived easiness and the perceived realism to see the objects in the virtual environment. The participants rated the horizontal ($M = 4.48$, $SD = 0.59$) lower than the vertical ($M = 4.60$, $SD = 0.58$) orientation. For both orientations, the perceived accuracy of visual cues was higher in flat virtual content than in curved. When individually comparing the scores for perceived realism of objects, there was a main effect of flat horizontal vs. curved horizontal ($p < 0.04$). flat models had the same mean, so there was also a main effect when comparing flat vertical vs. curved horizontal ($p < 0.04$).

- **Willingness to Recommend** The willingness to recommend scores were rated on a ten-point Likert scale. The mean score show the horizontal ($M = 8.98$, $SD = 1.38$) and vertical ($M = 8.98$, $SD = 1.48$).

## 4.3 User Feedback

When asked about their thoughts on improving the VR drawing activity for future use, participants responded with feedback on the difficulty, easiness of seeing, and the stretch required to reach the tasks. Participant #6 reported that for the curved configuration, in horizontal orientation, it was *"hard to reach the furthest part of the fish from standing in one spot"* and in the vertical orientation, it was *"hard to tell what dots were farther away/closer to me."* This shows how orientation affects how much a user perceives body stretch while reaching and estimating distance from an object in a VE. When drawing the flat fish in a vertical orientation, they expressed that *"this level was too easy."* Similarly, Participant #11 expressed that the flat model drawn vertically *"felt like I finished this model really fast so [it] wouldn't be that practical for a game."*

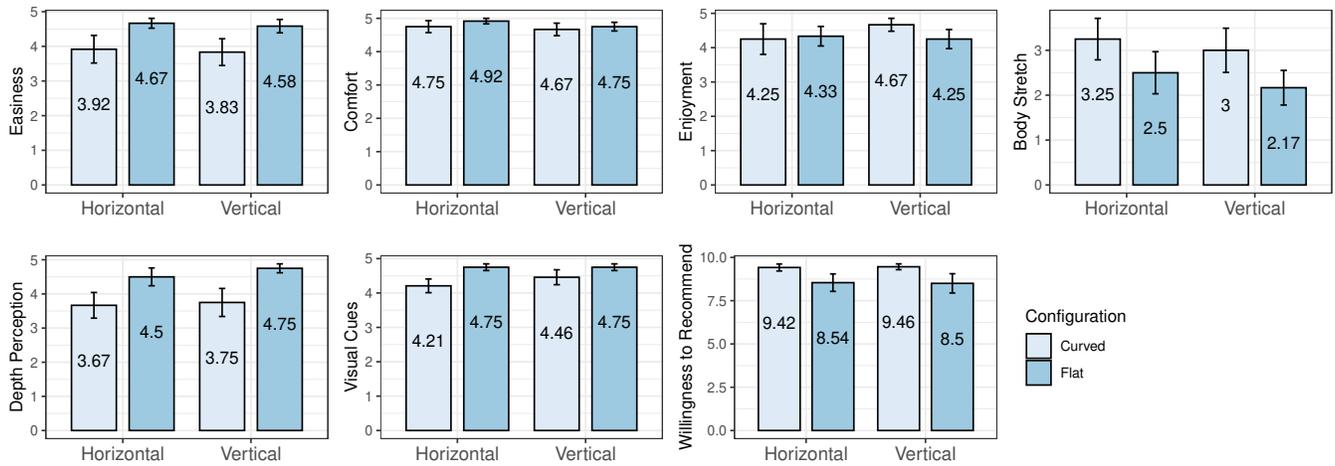

**Figure 6: Questionnaire results of the average perceived scores for easiness, comfort, enjoyment, body stretch, depth perception, visual cues, and willingness to recommend.**

Such suggestions offer insight into how our task can be used as short repeatable exercises for patients that are not too strenuous in each iteration.

Participant #12 said that they liked the flat model more in horizontal orientation than vertical orientation because they could *"look down on it as I draw"* and the horizontal task was *"realistic, easy, enjoyable."* However, Participant #9 expressed that with flat model in horizontal orientation, *"it was a little challenging to see the order of the dots in this setting, and I think I traced some of the dots out of order because of this."*

## 5 DISCUSSION

In the following section, we discuss the results from our preliminary user study with respect to the hypotheses described in Section 3.5.

***Effects of Configuration.*** The results show statistically significant differences in the variable *resistance change* for model configuration. The **H1-1** is not supported because the flat configuration induced more electrical resistance changes compared to the curved configuration, in particular, in the vertical orientation. It reveals the benefits of flat configuration, which induces more elbow stretches for rehabilitation.

There was no significant difference on the main effect between the configurations in term of task completion time. However, there is a significant difference on their interaction effect between the configuration and orientation. The results show that the flat configuration was performed faster in the vertical orientation. For the number of mistakes, there was no significant difference in the main effect of configurations. However, according to descriptive results, the average of mistakes for the flat configuration was lower than for the curved configuration. Based on the results, we state that the **H1-2** is supported.

***Effects of Orientation.*** The results of objective measures do not support the **H2-1**. In contrast, the vertical condition induced more electrical resistance changes compared to the horizontal condition. It is also noteworthy that there was a significant difference on the flat configuration between both orientations. The results indicate that the resistance change in the vertical orientation for flat configuration was greater than in the horizontal orientation.

For **H2-2**, the results of the normalized task completion time reveal that the vertical outperformed the horizontal orientation in the flat configuration. There was no significant difference in terms of the normalized number of mistakes. However, the average number of mistakes of vertical was lower than horizontal orientation. Therefore, we assume that the results support the **H2-2**.

The results of the objective measures for elbow flexion reveal that the participants' performance was influenced by the orientations and configurations of the VR therapy exergame. However, there were no statistically significant differences on the main effects of task completion and number of mistakes of both conditions. It indicates that the VR therapy exergame can lead to customizing therapy while maintaining the same intensity level. For example, if the patient has restricted lower limb and required to be seated, they can use different orientations and configurations interchangeably.

For *subjective measures*, the participants reported relatively higher scores for *willing to recommend* for the multi-dimensional *curved* content. This encourages researchers to look at how more curves and dimensions and the combination of reaching up/down/up/in movements can be used in future studies of enjoyable tele-rehabilitation.

***Limitations and future work.*** One of the major limitations was that the wearable sleeve was *one size fits all*. Thus, for some participants, it was too loose and needed to be secured with rubber bands. Besides, the signal for the sleeve sensor was sometimes disconnected because of the fragility of the hardware. Thus, more work for robust data collection with this sensor is needed, in addition to making different sizes of the elbow sleeve sensor [7]. Also, though the creative drawing model was meant to be drawn in one continuous arm stroke, some participants had a difficult time following our intended path. Future work will include guided paths so that there is more consistency in resistance change data from the same task. This is important because, in the future, we can look at distance misestimation for depth in terms of overestimation (controller goes past the dots) or underestimations (controller did not reach the dot), especially with respect to Fitts' Law [23] in a simpler virtual content. This will give us an idea of where depth perception falls short in our immersive virtual exergame.

In addition, there are more dimensions that can be further developed for VR therapy exergame. For example, implementing multiplayer feature allows collaboration among users, thereby making the exergame more interesting and improving user experience [6, 22]. Collecting multi-model data during experiment is also worth investigation. The collected data can be used to train machine learning models which predicts the exercise intensity. Hence, future research can utilize the collected data and trained model to recognize key features that have the biggest impact on the outcome, i.e. quantitative and qualitative assessments. Another limitation of our study was the small sample size of our participation pool and volunteer-based recruitment for convenience. From the perspective of improving on the experimental settings, future studies should co-operate with medical institutions and invite the target population of upper extremity therapy, such as Parkinson's patients, post-stroke patients or people who have experienced serious injuries.

## 6 CONCLUSION

In this paper, we presented a VR therapy exergame for upper extremity rehabilitation that captured both hand and elbow joint movement with a smart wearable sensor. The results provide insights that the orientation and configuration of virtual content may possibly be used for therapeutic applications. Such therapy does not decrease patient accuracy or depth perception and importantly may increase the movement of the involved joint. Our findings show that these study conditions may be appropriate for exercises that prefer a seated position of the user and want to focus on upper extremity mobility. Further research is necessary to study how to improve visual cues, depth perception, and reaching capabilities for 3D and curved objects in virtual environments, especially for VR therapy where accuracy is important for a patient's healthcare. Our results provide insights and show potential research directions for at-home virtual therapy, especially by using multi-model sensing data for further in-depth analysis of limb movement and range of motion assessment.

## REFERENCES
[1] Saba Anwer, Asim Waris, Syed Omer Gilani, Javaid Iqbal, Nusratnaaz Shaikh, Amit N Pujari, and Imran Khan Niazi. 2022. Rehabilitation of Upper Limb Motor Impairment in Stroke: A Narrative Review on the Prevalence, Risk Factors, and Economic Statistics of Stroke and State of the Art Therapies. In *Healthcare*, Vol. 10. 190.
[2] Paulina JM Bank, Marina A Cidota, P Elma W Ouwehand, and Stephan G Lukosch. 2018. Patient-tailored augmented reality games for assessing upper extremity motor impairments in Parkinson's disease and stroke. *Journal of medical systems* 42, 12 (2018), 1–11.
[3] Lauren Baron, Qile Wang, Sydney Segear, Brian A Cohn, Kangsoo Kim, and Roghayeh Barmaki. 2021. Enjoyable Physical Therapy Experience with Interactive Drawing Games in Immersive Virtual Reality. In *Symposium on Spatial User Interaction*. 1–8.
[4] Alexandre Fonseca Brandão, Diego Roberto Colombo Dias, Sávyo Toledo Machado Reis, Clovis Magri Cabreira, Maria Cecilia Moraes Frade, Thomas Beltrame, Marcelo de Paiva Guimarães, and Gabriela Castellano. 2020. Biomechanics sensor node for virtual reality: A wearable device applied to gait recovery for Neurofunctional rehabilitation. In *International Conference on Computational Science and Its Applications*. Springer, 757–770.
[5] Pablo Campo-Prieto, José María Cancela, and Gustavo Rodríguez-Fuentes. 2021. Immersive virtual reality as physical therapy in older adults: present or future (systematic review). *Virtual Reality* 25, 3 (2021), 801–817.
[6] Vuthea Chheang, Danny Schott, Patrick Saalfeld, Lukas Vradelis, Tobias Huber, Florentine Huettl, Hauke Lang, Bernhard Preim, and Christian Hansen. 2022. Towards Virtual Teaching Hospitals for Advanced Surgical Training. In *2022 IEEE Conference on Virtual Reality and 3D User Interfaces Abstracts and Workshops (VRW)*. 410–414.
[7] Sagar M Doshi, Colleen Murray, Amit Chaudhari, Dae Han Sung, and Erik T Thostenson. 2022. Ultrahigh sensitivity wearable sensors enabled by electrophoretic deposition of carbon nanostructured composites onto everyday fabrics. *Journal of Materials Chemistry C* 10, 5 (2022), 1617–1624.
[8] Franz Faul, Edgar Erdfelder, Albert-Georg Lang, and Axel Buchner. 2007. G* Power 3: A flexible statistical power analysis program for the social, behavioral, and biomedical sciences. *Behavior research methods* 39, 2 (2007), 175–191.
[9] Holly C. Gagnon, Taren Rohovit, Hunter Finney, Yu Zhao, John M. Franchak, Jeanine K. Stefanucci, Bobby Bodenheimer, and Sarah H. Creem-Regehr. 2021. The Effect of Feedback on Estimates of Reaching Ability in Virtual Reality. In *IEEE Virtual Reality and 3D User Interfaces (VR)*. 798–806.
[10] Nicolas Gerig, Johnathan Mayo, Kilian Baur, Frieder Wittmann, Robert Riener, and Peter Wolf. 2018. Missing depth cues in virtual reality limit performance and quality of three dimensional reaching movements. *PLoS one* 13, 1 (2018).
[11] Aaron J. Hartstein, Margaret Verkuyl, Kory Zimney, Jean Yockey, and Patti Berg-Poppe. 2022. Virtual Reality Instructional Design in Orthopedic Physical Therapy Education: A Mixed-Methods Usability Test. *Simulation & Gaming* 53, 2 (2022), 111–134.
[12] Amy Henderson, Nicol Korner-Bitensky, and Mindy Levin. 2007. Virtual reality in stroke rehabilitation: a systematic review of its effectiveness for upper limb motor recovery. *Topics in stroke rehabilitation* 14, 2 (2007), 52–61.
[13] Lewis A Ingram, Annie A Butler, Matthew A Brodie, Stephen R Lord, and Simon C Gandevia. 2021. Quantifying upper limb motor impairment in chronic stroke: A physiological profiling approach. *Journal of Applied Physiology* 131, 3 (2021), 949–965.
[14] Mateus Michelin Jurioli, Alexandre Fonseca Brandao, Bárbara Cristina Silva Guedes Martins, Eduardo do Valle Simões, and Cláudeo Fabino Motta Toledo. 2020. Wearable device for immersive virtual reality control and application in upper limbs motor rehabilitation. In *International Conference on Computational Science and Its Applications*. 741–756.
[15] A Donald Keedwell and József Dénes. 2015. *Latin squares and their applications*.
[16] Won-Seok Kim, Sungmin Cho, Jeonghun Ku, Yuhee Kim, Kiwon Lee, Han-Jeong Hwang, and Nam-Jong Paik. 2020. Clinical application of virtual reality for upper limb motor rehabilitation in stroke: review of technologies and clinical evidence. *Journal of clinical medicine* 9, 10 (2020), 3369.
[17] Efthimis Kioumourtzoglou, Vassiliki Derri, Olga Mertzanidou, and George Tzetzis. 1997. Experience with perceptual and motor skills in rhythmic gymnastics. *Perceptual and motor skills* 84, 3 (1997), 1363–1372.
[18] Peter Langhorne, Fiona Coupar, and Alex Pollock. 2009. Motor recovery after stroke: a systematic review. *The Lancet Neurology* 8, 8 (2009), 741–754.
[19] KE Laver, B Lange, S George, JE Deutsch, G Saposnik, and M Crotty. 2017. Virtual reality for stroke rehabilitation.
[20] Si-Huei Lee, Shih-Ching Yeh, Rai-Chi Chan, Shuya Chen, Geng Yang, and Li-Rong Zheng. 2016. Motor ingredients derived from a wearable sensor-based virtual reality system for frozen shoulder rehabilitation. *BioMed research international* 2016 (2016).
[21] Angela Li, Zorash Montaño, Vincent J Chen, and Jeffrey I Gold. 2011. Virtual reality and pain management: current trends and future directions. *Pain management* 1, 2 (2011), 147–157.
[22] Hui Liang, Shiqing Liu, Yi Wang, Junjun Pan, Yazhou Zhang, and Xiaohang Dong. 2023. Multi-user upper limb rehabilitation training system integrating social interaction. *Computers & Graphics* (2023).
[23] I Scott MacKenzie. 1992. Fitts' law as a research and design tool in human-computer interaction. *Human-computer interaction* 7, 1 (1992), 91–139.
[24] Andrea H Mason, Masuma A Walji, Elaine J Lee, and Christine L MacKenzie. 2001. Reaching movements to augmented and graphic objects in virtual environments. In *Proc. of the SIGCHI conference on Human factors in computing systems*. 426–433.
[25] Jaume Morera-Balaguer, José Martín Botella-Rico, Mari Carmen Martínez-González, Francesc Medina-Mirapeix, and Óscar Rodríguez-Nogueira. 2018. Physical therapists' perceptions and experiences about barriers and facilitators of therapeutic patient-centred relationships during outpatient rehabilitation: a qualitative study. *Brazilian journal of physical therapy* 22, 6 (2018), 484–492.
[26] Anzalna Narejo, Attiya Baqai, Neha Sikandar, Ali Absar, and Sanam Narejo. 2020. Physiotherapy: Design and Implementation of a Wearable Sleeve using IMU Sensor and VR to Measure Elbow Range of Motion. *International Journal of Advanced Computer Science and Applications* 11, 9 (2020).

[27] Adrian KT Ng, Leith KY Chan, and Henry YK Lau. 2016. Depth perception in virtual environment: The effects of immersive system and freedom of movement. In *International Conference on Virtual, Augmented and Mixed Reality*. 173–183.
[28] National Institute of Neurological Disorders and Stroke. 2008. Post-Stroke Rehabilitation Fact Sheet. NIH Publication 80-4846.
[29] American Academy of Orthopaedic Surgeons et al. 2016. One in two Americans have a musculoskeletal condition: New report outlines the prevalence, scope, cost and projected growth of musculoskeletal disorders in the US. *ScienceDaily. ScienceDaily* 1 (2016).
[30] Shanmugam Muruga Palaniappan, Shruthi Suresh, Jeffrey M Haddad, and Bradley S Duerstock. 2020. Adaptive Virtual Reality Exergame for Individualized Rehabilitation for Persons with Spinal Cord Injury. In *European Conference on Computer Vision*. Springer, 518–535.
[31] Jack Parker, Lauren Powell, Susan Mawson, et al. 2020. Effectiveness of upper limb wearable technology for improving activity and participation in adult stroke survivors: systematic review. *Journal of medical Internet research* 22, 1 (2020).
[32] Ivan Phelan, Penny Jayne Furness, Maria Matsangidou, Alicia Carrion-Plaza, Heather Dunn, Paul Dimitri, and Shirley A. Lindley. 2021. Playing your pain away: designing a virtual reality physical therapy for children with upper limb motor impairment. *Virtual Reality* (Jun 2021).
[33] Sandeep K Subramanian, MacKenzie K Cross, and Cole S Hirschhauser. 2022. Virtual reality interventions to enhance upper limb motor improvement after a stroke: commonly used types of platform and outcomes. *Disability and Rehabilitation: Assistive Technology* 17, 1 (2022), 107–115.
[34] Katy Tcha-Tokey, Olivier Christmann, Emilie Loup-Escande, and Simon Richir. 2016. Proposition and validation of a questionnaire to measure the user experience in immersive virtual environments. (2016).
[35] Robert Teasell, Matthew J Meyer, Andrew McClure, Cheng Pan, Manuel Murie-Fernandez, Norine Foley, and Katherine Salter. 2009. Stroke rehabilitation: an international perspective. *Topics in stroke rehabilitation* 16, 1 (2009), 44–56.
[36] Daria Tsoupikova, Nikolay S Stoykov, Molly Corrigan, Kelly Thielbar, Randy Vick, Yu Li, Kristen Triandafilou, Fabian Preuss, and Derek Kamper. 2015. Virtual immersion for post-stroke hand rehabilitation therapy. *Annals of biomedical engineering* 43, 2 (2015), 467–477.
[37] Andrea Turolla, Mauro Dam, Laura Ventura, Paolo Tonin, Michela Agostini, Carla Zucconi, Pawel Kiper, Annachiara Cagnin, and Lamberto Piron. 2013. Virtual reality for the rehabilitation of the upper limb motor function after stroke: a prospective controlled trial. *Journal of neuroengineering and rehabilitation* 10, 1 (2013), 1–9.
[38] Sebastian Wagner, Fabian Joeres, Mareike Gabele, Christian Hansen, Bernhard Preim, and Patrick Saalfeld. 2019. Difficulty factors for VR cognitive rehabilitation training–Crossing a virtual road. *Computers & Graphics* 83 (2019), 11–22.
[39] Frieder Wittmann, Jeremia P Held, Olivier Lambercy, Michelle L Starkey, Armin Curt, Raphael Höver, Roger Gassert, Andreas R Luft, and Roman R Gonzenbach. 2016. Self-directed arm therapy at home after stroke with a sensor-based virtual reality training system. *Journal of neuroengineering and rehabilitation* 13, 1 (2016), 1–10.
[40] Yangfan Xu, Meiqinzi Tong, Wai-Kit Ming, Yangyang Lin, Wangxiang Mai, Weixin Huang, and Zhuoming Chen. 2021. A Depth Camera–Based, Task-Specific Virtual Reality Rehabilitation Game for Patients With Stroke: Pilot Usability Study. *JMIR Serious Games* 9, 1 (24 Mar 2021).
[41] Kevin Yu, Roghayeh Barmaki, Mathias Unberath, Albert Mears, Joseph Brey, Tae Hwan Chung, and Nassir Navab. 2018. On the accuracy of low-cost motion capture systems for range of motion measurements. In *Medical Imaging 2018: Imaging Informatics for Healthcare, Research, and Applications*, Vol. 10579. SPIE, 90–95.